# A short note on the two most recent large EQs of New Zealand (Mw=7.0 on September 3rd, 2010 and Mw=6.1 on February 21st, 2011). A typical example of tidally triggered large EQs by the M1 tidal component.

**Thanassoulas[1], C., Klentos[2], V., Verveniotis, G.[3], Zymaris, N.[4]**

1. Retired from the Institute for Geology and Mineral Exploration (IGME), Geophysical Department, Athens, Greece.
   e-mail: thandin@otenet.gr - URL: www.earthquakeprediction.gr

2. Athens Water Supply & Sewerage Company (EYDAP),
   e-mail: klenvas@mycosmos.gr - URL: www.earthquakeprediction.gr

3. Ass. Director, Physics Teacher at 2nd Senior High School of Pyrgos, Greece.
   e-mail: gver36@otenet.gr - URL: www.earthquakeprediction.gr

4. Retired, Electronic Engineer.

**Abstract**

The time of occurrence of the two most resent large EQs of New Zealand (Mw = 7.0 on September 3rd, 2010 and Mw = 6.1 on February 21st, 2011) is compared to the time when the lithospheric oscillation at the focal area reaches its peak amplitude in terms of the M1 (Moon declination) tidal component. The observed very good coincidence of the EQs occurrence time with the lithospheric oscillation peak amplitude time for the M1 tidal component clearly suggests that both these EQs were triggered by the M1 component of the lithospheric tidal oscillation. It is speculated that large preseismic electric signals should be observed before the EQs occurrence if the appropriate hardware were installed.

**Key words:** New Zealand, large earthquakes, M1 tidal wave, lithospheric oscillations, short-term earthquake prediction.

## 1. Introduction.

Although the large earthquakes are generally considered by the majority of the seismological community, as unpredictable in time, in terms of statistics, there exist well-known physical mechanisms which trigger their occurrence on specific favorable times. Such a typical mechanism is the presence of the lithospheric tidal waves which are generated by the Sun – Moon and Earth system gravity interaction. Papers on this topic, the triggering of EQs by tidal waves, have been referred extensively by Thanassoulas (2007) while a detailed account of the triggering mechanism and its way for generating extensional or compressional stress-loads in the lithosphere has been presented in the same reference too.

In brief, the following figure (1) shows the way the lithospheric tidal waves affect a seismogenic area. In the simplest case of the seismogenic model (fig. 1. left), the stress-load of the seismogenic area increases very slowly and linearly in time (inclined blue line), due to long term tectonic changes, until it reaches the rock failure stress load level of the focal area that is represented by the horizontal red line and then the EQ occurs (red arrow). Typically, the time of occurrence of the seismogenic area rock failure (EQ), represented by the red arrow is undefined since nor the rock failure stress load level neither the long term stress load increase rate are known in advance.

Next, the adopted model will be improved by adding the lithospheric tidal M1 (Moon declination) oscillatory component. The way the stress load increase rate in the seismogenic area is modified and is presented in figure (1, right). It is pretty clear that the rock failure stress load level of the seismogenic area will be reached by the total stress load resulting from the addition of the long range linear stress load plus the oscillating tidal stress load component caused by the oscillating lithospheric deformation (see details in Thanassoulas, 2007). Consequently, the EQ will not occur at any random time but on the amplitude peak time of the lithospheric tidal wave that firstly reaches the rock failure stress-load level.

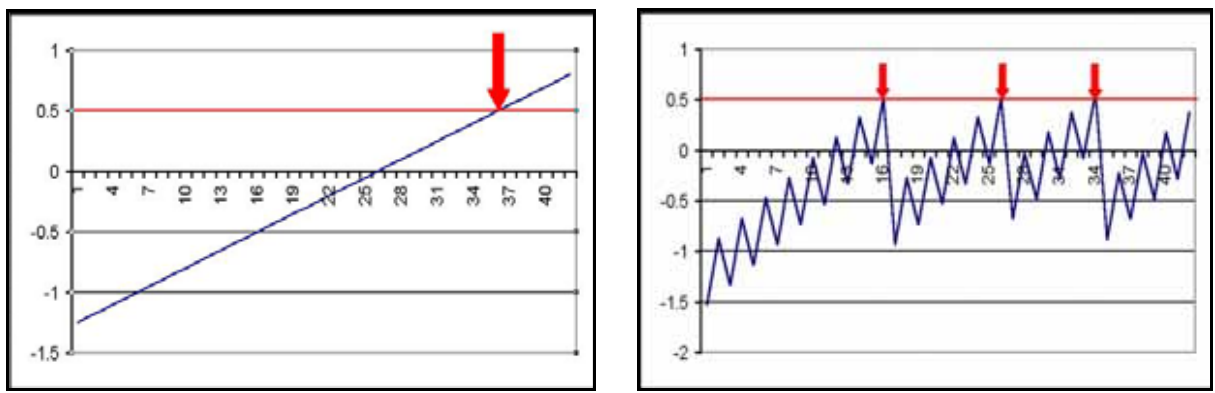

Fig. 1. Left = Occurrence of a large EQ (red arrow) considering a simple seismogenic model. Right = Occurrence of large EQs considering lithospheric tidal oscillation too. The red horizontal line represents the seismogenic area rock failure stress-load level (in arbitrary units) while the blue line represents the stress load at the seismogenic area as a time function.

The briefly presented theoretical analysis will be tested against the case of the most recent New Zealand large EQs, of the September 3$^{rd}$, 2010 earthquake of a magnitude of Mw = 7.0 and the February 21$^{st}$, 2011 earthquake of a magnitude of Mw = 6.1 (magnitudes as determined by EMSC).

The lithospheric tidal oscillation will be calculated by the Rudmam et al. (1977) method at a minute's sampling interval while the M1 component will be determined by the appropriate low-pass filtering in order to reject oscillatory tidal components of periods shorter than this of the M1 (T = 14 days).

**2. The most recent large EQs of New Zealand.**

The two most recent large New Zealand EQs will be presented as: 1) location maps compiled by the EMSC accompanied by 2) the corresponding compiled graph that indicates the tidal M1 component oscillation and consequently the corresponding lithospheric one too. The M1 component has been calculated for the following geographical coordinates: Lat. = -43.5, Lon. = 172.0 for a period of almost one month before and following after each main seismic event.

**2.1 EQ on 3$^{rd}$ of September, 2010 with a magnitude of Mw = 7.0**

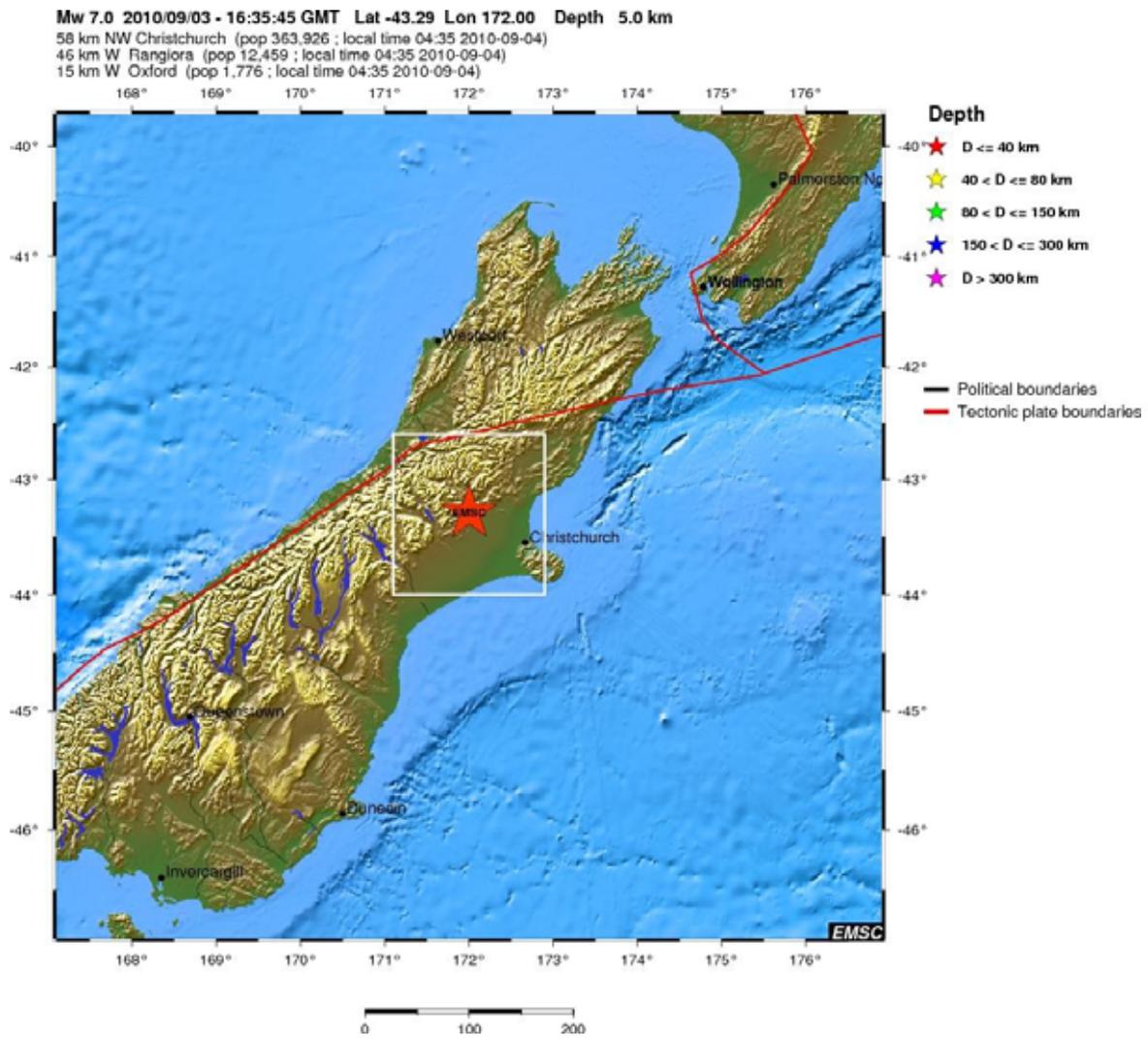

Fig. 2. Location map (after EMSC) of the EQ (red asterisk) of 3$^{rd}$ of September, 2010 with a magnitude of Mw = 7.0

**The corresponding tidal data.**

Next, the tidal oscillating M1 component will be compared to the time of occurrence of the corresponding large EQ. The comparison is presented in the following figure (3).

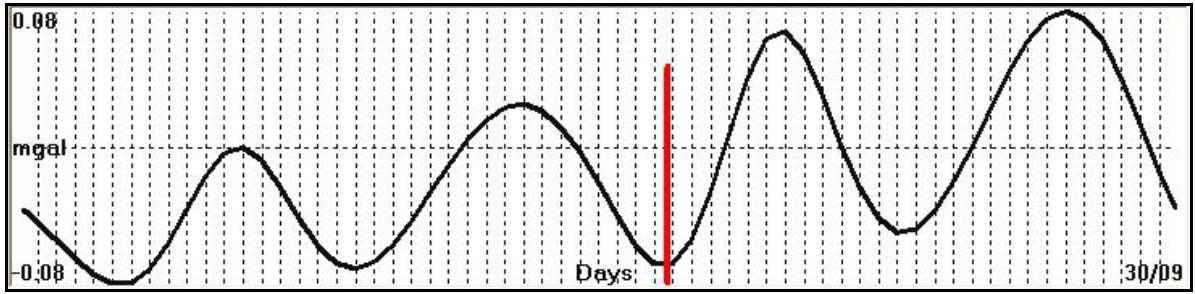

**Fig. 3.** Comparison of the M1 tidal oscillation (black line) with the time of occurrence (red bar) of the EQ of 3rd of September, 2010 (Mw = 7.0). Vertical scale is in mgals. Consequently, the lithosphere oscillates in the same mode.

In this case the EQ occurred, compared to the lithospheric tidal oscillation, exactly on the peak of the M1 tidal wave, as it was predicted by the physical mechanism presented in paragraph (1).

## 2.2. EQ of 21st of February, 2011 with a magnitude of Mw = 6.1

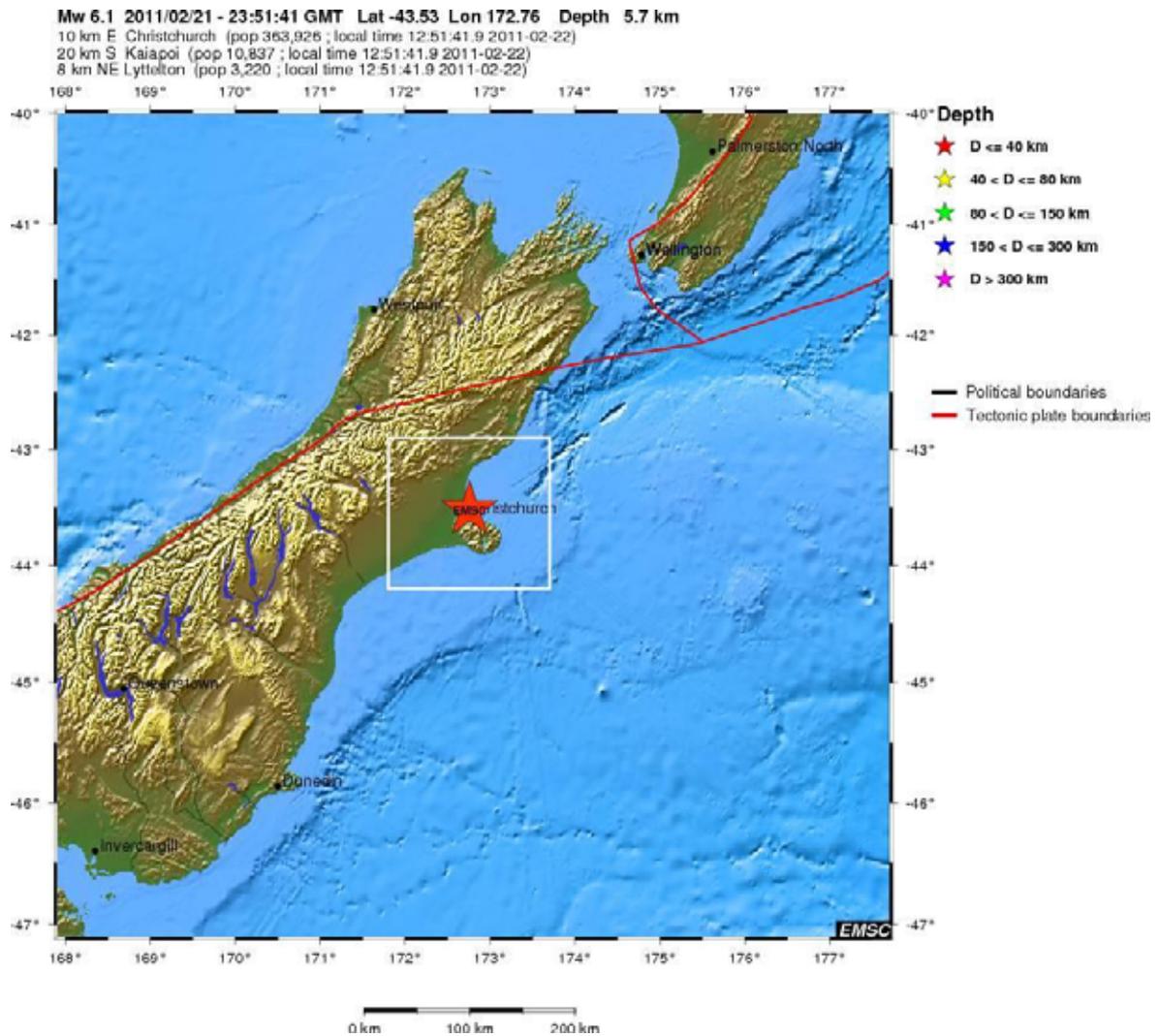

**Fig. 4.** Location map (after EMSC) of the EQ of 21st of February, 2011 with a magnitude of Mw = 6.1

**The corresponding tidal data.**

Next, the lithospheric oscillating M1 component will be compared to the time of occurrence of the corresponding large EQ. The comparison is presented in the following figure (5).

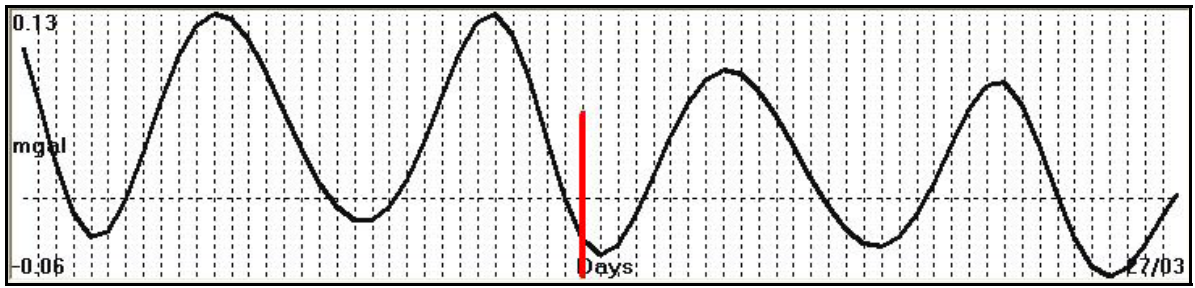

**Fig. 5.** Comparison of the M1 tidal oscillation (black line) with the time of occurrence (red bar) of the EQ of 21$^{st}$ of February, 2011 with a magnitude of Mw = 6.1. Vertical scale is in mgals. Consequently, the lithosphere oscillates in the same mode.

The specific EQ deviated for only one (1) day from the corresponding lithospheric oscillating tidal peak of the M1 component.

## 3. Conclusions.

In an earlier study (Thanassoulas, 2007), performed on data obtained from the Greek seismogenic territory, from the analysis of forty (40) EQs with magnitude of Ms = > 5.5R the following results were obtained:

a. By chance ($P_{ch}$) coincidence of the tidal peak time with the EQ occurrence time within an accuracy of +/- a day equals to:

$$P_{ch} = 14.28\% \text{ for a half period (7 days)}$$

b. **Mean value** of time difference of the time of occurrence of the EQs, from the corresponding tidal peak times, for the entire data set (40 EQs), equals to 1.18 days.

$$\text{dt Mean Value} = 1.18 \text{ days}$$
$$\text{S. Dev. Of dt} = 1.15 \text{ days}$$

Percentage of EQs with dt = 0    days equals to    39.47% (exact day)

Percentage of EQs with dt = 0 - 1 days equals to    50.00%

Percentage of EQs with dt = 0 - 2 days equals to    78.95%

Percentage of EQs with dt = 0 - 3 days equals to 100.00 %

Considering the time deviations of dt = 0 days and dt = 0-1 days, in other words assuming quite short-term time prediction, it is shown that the earlier presented large EQs triggering mechanism behaves by far much better than a random coincidence.

Both the recent New Zealand large EQs performed within the frame of the results of the former study. The first one (2010) coincided exactly with the tidal peak time, therefore it is placed in the group that is characterized by 39.47% success rate, while the second one (2011) deviates for only one day, therefore it is placed in the group that is characterized by 50.00%.

Consequently, both EQs did not occur randomly in time. On the contrary, these EQs were clearly triggered by the lithospheric tidal oscillatory M1 component.

Finally, we believe, according to our experience obtained from a current long lasting experiment on preseismic electric signals detection conducted in Greece, that both these large EQs probably had generated a strong preseismic electric signal (due to the large lithospheric crystalline deformation that precedes the mechanical rapture of the focal area) which could be registered quite easily provided that the appropriate network hardware had been installed well before the EQs occurrence (see current research program and similar examples from large EQs in Greece at www.earthquakeprediction.gr ).

## 4. References.